\documentclass[10pt,journal,compsoc]{IEEEtran}

\usepackage[style=ieee,backend=biber,defernumbers=true,sorting=nyt]{biblatex}

\usepackage{caption}
\usepackage{subcaption}
\usepackage{graphicx}
\usepackage{amsmath}
\usepackage{multirow}
\usepackage{caption}

\usepackage[dvipsnames]{xcolor}
\usepackage{url}
\usepackage{booktabs}
\usepackage{multirow}
\usepackage{xspace}
\usepackage{float}
\usepackage{tcolorbox}
\usepackage{balance}
\usepackage{listings}
\lstdefinelanguage{JavaScript}{%
	keywords={const, let, typeof, instanceof, new, true, false, catch, function, return, null, undefined, catch, switch, var, if, in, while, for, do, else, case, break},
	keywordstyle=\bfseries,
	ndkeywords={class, export, throw, import, this},
	ndkeywordstyle=\bfseries,
    sensitive=false,
    comment=[l]{//},
	morecomment=[s]{/*}{*/},
	commentstyle=\ttfamily,
	stringstyle=\color{blue}\ttfamily,
}
\newcommand{\canvas}{\texttt{<canvas>}\xspace}

\makeatletter
\let\blx@rerun@biber\relax
\makeatother

\begin{document}

	\title{A Taxonomy of Testable HTML5 Canvas Issues}
	
	\author{Finlay~Macklon,
		Markos~Viggiato,		
		Natalia~Romanova,
		Chris~Buzon,
		Dale~Paas,
		Cor-Paul~Bezemer%
	\IEEEcompsocitemizethanks{
		\IEEEcompsocthanksitem Finlay Macklon, Markos Viggiato, and Cor-Paul Bezemer are with the Analytics of Software, Games and Repository Data (ASGAARD) lab at the University of Alberta, Canada.
		E-mail:\{macklon, viggiato, bezemer\}@ualberta.ca
		
		\IEEEcompsocthanksitem Natalia Romanova,  Chris Buzon, and Dale Paas are with Prodigy Education, Toronto, Canada.
		E-mail: \{natalia.romanova, christopher.buzon, dale.paas\}@prodigygame.com
	}}


	\date{January 14, 2022}
	
	\IEEEtitleabstractindextext{%
		\begin{abstract}
The HTML5 \canvas is widely used to display high quality graphics in web applications.
However, the combination of web, GUI, and visual techniques that are required to build \canvas applications, together with the lack of testing and debugging tools, makes developing such applications very challenging.
To help direct future research on testing \canvas applications, in this paper we present a taxonomy of  testable \canvas issues.
First, we extracted 2,403 \canvas-related issue reports from 123 open source GitHub projects that use the HTML5 \canvas.
Second, we constructed our taxonomy by manually classifying a random sample of 332 issue reports.
Our manual classification identified five broad categories of testable \canvas  issues, such as \textit{Visual} and \textit{Performance}  issues.
We found that \textit{Visual}  issues are the most frequent (35\%), while \textit{Performance}  issues are relatively infrequent (5\%). 
We also found that many testable \canvas  issues that present themselves visually on the \canvas are actually caused by other components of the web application.
Our taxonomy of testable \canvas issues can be used to steer future research into \canvas issues and testing.
		\end{abstract}
		
		\begin{IEEEkeywords}
				html5 canvas, web applications, issue taxonomy, issue reports
		\end{IEEEkeywords}
	}

	\maketitle

	\IEEEdisplaynontitleabstractindextext
	
	\IEEEpeerreviewmaketitle

\section{Introduction}
The HTML5 \canvas allows the rendering of high quality graphics in web applications.
The Canvas API and WebGL API each provide JavaScript methods for drawing graphics on the HTML \canvas element.
The \canvas is widely-used to develop web applications such as data visualizations, animations, and web games.
The use of \canvas in web applications is expected to grow even further, because its main alternative, Adobe Flash, is no longer supported in modern browsers as of 2021~\cite{flasheol, adobeblog}.

However, developers face challenges when working with the \canvas.
For example, a study of web development Q\&A forum posts found that the \canvas can cause confusion for developers, which is attributed to a lack of API documentation~\cite{bajaj2014mining}.
In terms of web applications, API documentation is arguably particularly important for the \canvas, as \canvas applications operate differently from traditional web applications. 
Traditional web applications are controlled through the Document Object Model (DOM), which represents the web page as nodes and objects~\cite{mozilladom}.
Testing methods for traditional web applications often rely on the DOM to test the page.
However, instead of being represented in the DOM, \canvas contents are represented as a bitmap which is directly manipulated using the Canvas API or WebGL API~\cite{mozillacanvas}.
The lack of a DOM representation of \canvas contents makes it less intuitive and much harder to build and test \canvas applications. 

To better understand the issues that developers face with the \canvas, we address the following research question in our study: \textit{What~types~of~~testable~issues~do~developers~encounter~when~creating~web~applications~with~the~HTML5~\canvas?}

In this paper, we construct a taxonomy of testable \canvas  issues to give better insights on how \canvas  issues are different from generic web application  issues, and to direct future research on testing the HTML5 \canvas. 
In our study, we build an understanding of the types of testable \canvas issues by manually analyzing the contents of 332 \canvas-related  issue reports in 123 open-source GitHub projects.
We combine an existing taxonomy of GUI  issues~\cite{lelli2015classifying} and an existing taxonomy of web issues~\cite{marchetto2009empirical} to provide a baseline taxonomy that contains an initial set of issue types.
Then, we perform a manual classification process to classify  issues into the existing types and define new  issue types as necessary.
The main contribution of our paper is a taxonomy of testable HTML5 \canvas  issues, which allows future researchers to better target their efforts on testing methods for the \canvas.

The remainder of our paper is structured as follows.
In Section~\ref{sec:background}, we provide background information to motivate our study.
In Section~\ref{sec:related}, we discuss related work.
We present our methodology in Section~\ref{sec:methodology}.
We present our taxonomy of testable \canvas  issues in Section~\ref{sec:ourtaxonomy}.
In Section~\ref{sec:discussion}, we compare \canvas  issues to generic web  issues and GUI  issues.
In Section~\ref{sec:futuredirections}, we discuss future research directions.
In Section~\ref{sec:threats}, we discuss threats to validity.
Section~\ref{sec:conclusion} concludes our paper.
	
\section{Background and Motivation} \label{sec:background}

\subsection{HTML5 \canvas applications}
The HTML5 \canvas is particularly useful for web applications which require dynamic graphics~\cite{operacanvas}, such as animations, interactive data visualizations, or web games.
A \canvas application can be made interactive by making it listen and respond to browser events such as mouse clicks~\cite{ibmcanvas}. 
After the occurrence of such an event, its associated JavaScript code is executed, which changes the \canvas bitmap.
This interactive behaviour allows developers to create `complete' applications inside the \canvas element.
As a result, many \canvas applications have their own GUI, which is rendered on the \canvas bitmap as well.
Because \canvas applications have a GUI, deal with graphics rendering, and execute inside of a web browser, a diverse skill set is needed to build such applications, making development challenging.

\subsection{Difficulties in \canvas testing} \label{sec:challenges}
\begin{figure}[t]
	\centering
	\begin{subfigure}{0.2\textwidth}
		\centering
		\includegraphics[width=\textwidth]{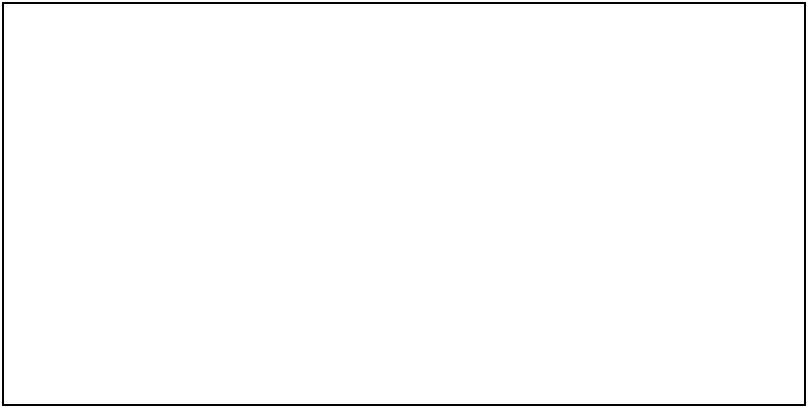}
		\caption{Before drawing.}
		\label{fig:before}
	\end{subfigure}
	\hfill
	\begin{subfigure}{0.2\textwidth}
		\centering
		\includegraphics[width=\textwidth]{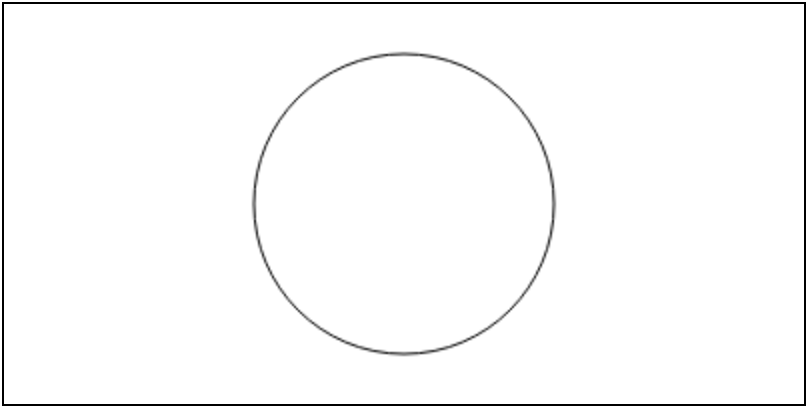}
		\caption{After drawing.}
		\label{fig:after}
	\end{subfigure}
	\par\bigskip
	\begin{subfigure}{0.4\textwidth}
		\centering
		\includegraphics[width=\textwidth]{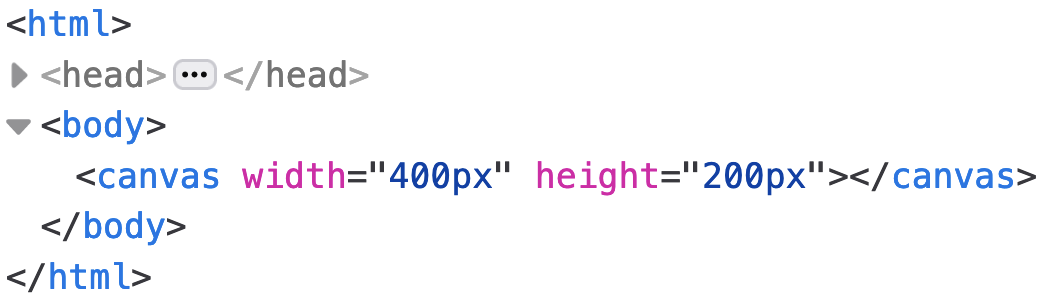}
		\caption{DOM both before and after drawing.}
		\label{fig:dom}
	\end{subfigure}
	\caption{\canvas contents are not represented in the DOM.}
	\label{fig:figures}
\end{figure}

To illustrate the challenges of testing the HTML5 \canvas, we provide a simple motivating example.
Figure \ref{fig:before} shows an empty \canvas element on a web page.
Figure \ref{fig:after} shows the updated \canvas element after executing the JavaScript code shown in Listing \ref{list:canvasapi}.
Figure \ref{fig:dom} shows the DOM representation of the \canvas element for both Figure \ref{fig:before} and Figure \ref{fig:after}.
Clearly, the \canvas contents have changed, but the DOM representation of the \canvas element has not changed, showing that the \canvas contents are not represented in the DOM.

\begin{figure}[t]
	\centering
	\begin{lstlisting}[caption={Using the Canvas API to draw a circle~\cite{mozillacanvasarc}.}, label={list:canvasapi},  language=JavaScript]
const canvas = document.querySelector('canvas'); 
const ctx = canvas.getContext('2d'); 
ctx.beginPath();
ctx.arc(200, 100, 75, 0, 2*Math.PI);
ctx.stroke();
	\end{lstlisting}
\end{figure}

As web testing is typically done through several widely-used test automation frameworks that rely on analyzing the DOM, such as \texttt{Selenium}\footnote{\label{selenium}\url{https://www.selenium.dev/}}, \texttt{Cypress}\footnote{\label{cypress}\url{https://www.cypress.io/}}, and \texttt{Playwright}\footnote{\label{playwright}\url{https://playwright.dev/}}, existing web testing techniques are severely limited in \canvas testing.
There are many websites which use the \canvas~\cite{similartech} yet there is no good way to test the \canvas, meaning these websites either use manual testing, or do not test the \canvas at all.
With a lack of available \canvas testing tools, there are opportunities for future research.
A  testable issue  taxonomy provides a detailed classification of possible issue  types, and a detailed classification is necessary to build testing tools that target different types of  testable issues ~\cite{zaman2011security}. Additionally, challenges in brainstorming useful approaches to testing can be mitigated by using a testable issue  taxonomy~\cite{vijayaraghavan2003bug}. A testable issue taxonomy can help identify gaps in our knowledge and direct future research~\cite{ralph2018toward}. Therefore, our taxonomy of testable \canvas  issues  can be used to guide future research on \canvas testing.
	
\section{Related Work} \label{sec:related}

\subsection{HTML5 \canvas  issues }
While there is extensive prior work on web  issues  and web testing, there is limited prior work that investigates \canvas  issues.
As a discussion of related work on non-\canvas web testing is outside the scope of our paper, we refer to the systematic literature review by Do{\u{g}}an et al.~\cite{dougan2014web} and the grey literature review by Ricca and Stocco~\cite{ricca2021web} for an overview of related work about web testing.

Bajammal and Mesbah~\cite{bajammal2018web} propose an approach for automated visual testing of the \canvas, and report high accuracy.
However, their approach is evaluated by injecting only a single type of testable \canvas  issue.
In our work, we investigate the types of testable \canvas  issues that developers encounter, so that future research on \canvas testing can target a wider range of relevant testable issues.

Hoetzlein~\cite{hoetzlein2012graphics} conducted a study on graphics performance in web applications, finding HTML5 \canvas showed different performance across different browsers.
However, a difference in performance may not constitute a  testable issue, and our study is concerned with many types of \canvas  issues, rather than just performance  issues on the \canvas.

Bajaj et al.~\cite{bajaj2014mining} investigated Stack Overflow questions asked by web developers, and found that developers faced implementation issues with the \canvas.
They reason that this is due to a lack of clear API documentation.
While this does suggest that developers have problems with the \canvas, our work further investigates problems with the \canvas by identifying the types of  testable issues  developers face.

\subsection{Related  issue  taxonomies}
There are some existing issue taxonomies that are related to the GUI, web, and graphics rendering characteristics of the \canvas.
Lelli et al.~\cite{lelli2015classifying} constructed a taxonomy of GUI faults that allows for the evaluation of GUI testing tools.
Marchetto et al.~\cite{marchetto2009empirical} constructed a taxonomy of web faults that focuses on web architecture faults.
Our work differs in that we focus specifically on testable \canvas issues, allowing us to determine whether these taxonomies are useful in describing \canvas issue types, and define new  issue  types for the \canvas as required.

Ocariza et al.~\cite{ocariza2016study} constructed a taxonomy of client-side JavaScript bugs, finding that most client-side JavaScript bugs are DOM-related.
As explained in Section \ref{sec:background}, the \canvas contents are not represented in the DOM, meaning the taxonomy of client-side JavaScript bugs has limited relevance to testable \canvas  issues.

Woo et al.~\cite{woo1996s} provide a taxonomy of problems in rendering algorithms.
While their taxonomy may be complimentary to our taxonomy (in describing the causes of \textit{Rendering}  issues), our work takes a higher-level and broader view of testable \canvas  issue types.

\section{Methodology} \label{sec:methodology}
\begin{figure}[t]
	\centering
	\includegraphics[width=\columnwidth]{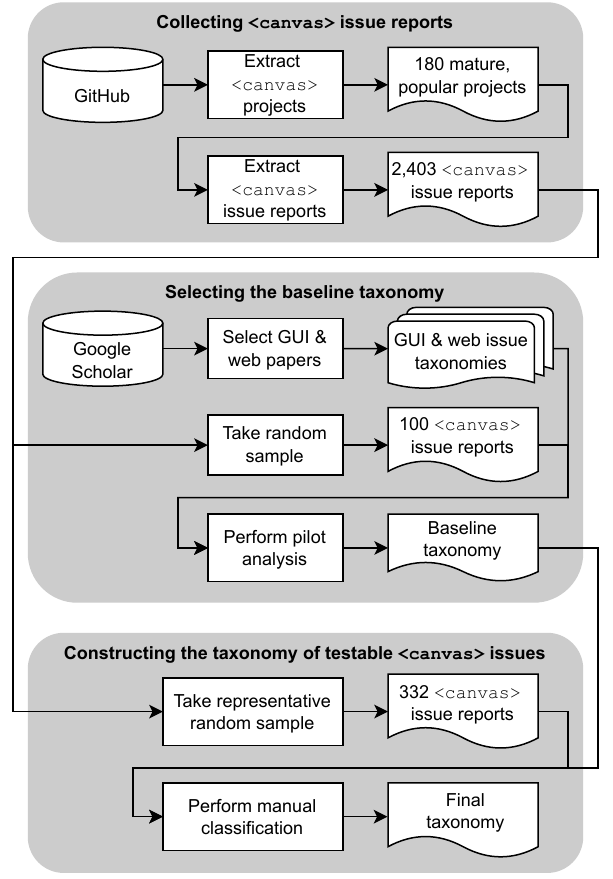}
	\caption{Study methodology overview.}
	\label{fig:methodology}
\end{figure}

In this section, we describe our methodology for constructing a taxonomy of testable \canvas  issues.
Table \ref{tab:guidelines} provides an overview of how we used taxonomy construction guidelines provided by Usman et al.~\cite{usman2017taxonomies} and Ralph~\cite{ralph2018toward} to facilitate the construction of our taxonomy. Issue taxonomies have been defined in a variety of ways, which is mainly due to differences in objectives~\cite{vijayaraghavan2003bug}. However, issue taxonomies have commonly been constructed via the manual classification of relevant issue reports~\cite{catolino2019not}. Our goal was to guide future research on \canvas testing by uncovering the types of testable issues (based on their symptoms) that are reported in \canvas projects, and manually classifying  issue reports provides the insight we require to build a taxonomy of testable \canvas  issues. Therefore to construct our taxonomy of testable \canvas  issues, we collected and manually classified \canvas  issue reports from GitHub. We took an initial set of labels from existing taxonomies, based on known aspects of the HTML5 \canvas: the \canvas is used as a GUI, for graphics rendering, and is a web technology.
An overview of our methodology can be seen in Figure \ref{fig:methodology}.

\begin{table}[t]
	\centering
	\caption{Mapping of taxonomy construction guidelines~\cite{usman2017taxonomies, ralph2018toward} to our methodology.}
	\begin{tabular*}{\linewidth}{l @{\extracolsep{\fill}} l @{\extracolsep{\fill}} l}
		\toprule
		\textbf{Usman et al.~\cite{usman2017taxonomies}} & \textbf{Ralph~\cite{ralph2018toward}} & \textbf{Section(s) of our paper} \\
		\midrule
		Planning & Choose a strategy & \ref{sec:methodology} \\
		\midrule
		Identification  & Site selection & \ref{sec:collecting} and \ref{sec:baseline} \\
		and extraction  & Data collection & \ref{sec:projects}, \ref{sec:reports}, and \ref{sec:selectingpapers} \\
		\midrule
		Design and & Data analysis & \ref{sec:pilot}, \ref{sec:construction}, \ref{sec:multilabel}, and \ref{sec:cardsorting} \\
		construction & & \\
		\midrule
		Validation & Conceptual evaluation & \ref{sec:evaluation} \\
		\bottomrule
	\end{tabular*}
	\label{tab:guidelines}
\end{table}

\subsection{Collecting \canvas  issue reports}\label{sec:collecting}
To analyze the types of  issues that developers face when using the \canvas, we looked at \canvas-related  issue reports in popular open source projects that use the \canvas. Issue reports in popular open-source projects that use the \canvas are appropriate to study because they can contain detailed descriptions and discussions of real-world  testable issues encountered by software developers, as shown by previous studies that mine  issue reports in popular open source projects~\cite{zaman2011security, catolino2019not, anvik2005coping, anvik2006should, anvik2007determining}.
We have made our collected data available on Zenodo at the following link: \url{https://zenodo.org/record/6886143}

\subsubsection{Extracting \canvas projects} \label{sec:projects}
We created a custom crawler that utilized the GitHub API to search GitHub for open source projects that use the \canvas.
We searched for open source projects on GitHub that matched the keywords \textit{`canvas'}, \textit{`html-canvas'}, or \textit{`html5-canvas'} in their title, description, readme file, or topic tags.
We limited our search to projects which use the following languages: HTML, JavaScript, or TypeScript.
To ensure only mature, popular projects were considered in this study, inclusion criteria of at least 1,000 stars and at least 2 contributors were set empirically, similar to prior work~\cite{bogner2022type, leitner2017exploratory, kamienski2021pysstubs}. We avoid immature projects because  issue reports for immature projects are not necessarily representative of real  testable issues, for example because the report is simply for a feature that is not finished or well-tested yet. Focusing on mature projects should improve the likelihood of  issue reports being more representative.
With our inclusion criteria applied, we had a list of 375 open source projects that matched our search criteria.
We then manually filtered out false positives.
Some false positives were due to our use of keyword matching in project readme files.
For example, one false positive was \texttt{microsoft/Web-Dev-For-Beginners}, which contains some guides for the \canvas but does not utilize it in a specific software application.
Other false positives were projects which utilized a non-HTML5 canvas object, such as \texttt{node-canvas} for Canvas API compatibility with \texttt{Node.js} desktop applications.
For projects with a non-HTML5 canvas object, it was difficult to automatically ascertain whether bug reports discuss the \canvas of interest, or the other canvas object.
With the false positives removed, we were left with a list of 180 projects from which \canvas-related  issue reports could be extracted.

\subsubsection{Extracting \canvas issue reports} \label{sec:reports}
To collect  issue reports for our analysis, we modified our custom crawler to search GitHub for relevant issue reports in each of the 180 projects.
To ensure we study only relevant  issue reports, we only include closed  issue reports in our analysis, similar to prior work~\cite{catolino2019not, tan2014bug}. Closed  issue reports contain the information required to understand the main symptom of the reported testable issue~\cite{catolino2019not}, for example, developer responses that confirm that the reported issue was indeed a testable issue and other relevant discussion in the comments of the  issue report.
We searched for closed issue reports with descriptions that matched the keyword \textit{`canvas'}.
Our search returned a list of 9,672 closed issue reports, from which we were able to identify testable issue reports.
For GitHub issue reports that were labelled, we selected issue reports that had a label containing the keyword \textit{`bug'} (except for labels like \textit{`not bug'}).
For GitHub issue reports that were not labelled, we identified testable issue reports using a keyword-based search on the issue report descriptions, similar to prior work~\cite{catolino2019not, karampatsis2020often, tan2014bug, kamienski2021pysstubs}.
Specifically, we selected issue reports with descriptions that matched one of the following keywords: \textit{`error'}, \textit{`bug'}, \textit{`fix'}, or \textit{`fault'}.
Our selection process provided 1,105 labelled issue reports and 1,301 unlabelled issue reports.
To ensure specific  issue reports were not over-represented in our population of collected  issue reports, we removed  issue reports that had been marked as duplicates on GitHub by the repository maintainer(s).
After removing  issue reports that were marked as duplicates on GitHub, we were left with a total of 2,403 \canvas-related issue reports.
We only identified relevant  issue reports in 123 of the projects.

\subsection{Selecting the baseline taxonomy\label{sec:baseline}}
To reduce subjectivity during the manual issue classification, we started our classification from a baseline taxonomy.
To select the baseline taxonomy, a pilot analysis was conducted with a taxonomy of GUI faults~\cite{lelli2015classifying} and a taxonomy of web faults~\cite{marchetto2009empirical}.

\subsubsection{Selecting papers with relevant taxonomies} \label{sec:selectingpapers}
As a web technology, we expected the \canvas to exhibit  testable issues that are present in web applications.
Also, the \canvas is often used for both the GUI and graphics of web applications. 
Therefore, we also expected the \canvas to exhibit  testable issues that are present in the GUIs of desktop applications.
To find relevant issue taxonomies, we searched Google Scholar with the following queries: \textit{`web bug taxonomy'} and \textit{`gui bug taxonomy'}.
The use of Google Scholar to find relevant software engineering works is suggested by Brereton et al.~\cite{brereton2007lessons}, and provides a wide coverage of search results as discussed by Landman et al.~\cite{landman2017challenges}.
The search results were manually analyzed, and two papers providing a relevant taxonomy were selected (one for GUI issues~\cite{lelli2015classifying} and one for web issues~\cite{marchetto2009empirical}).

\subsubsection{Performing the pilot analysis} \label{sec:pilot}
A pilot analysis was conducted by the first author to verify that each of the selected taxonomies were useful for our purposes.
Manual classification was performed with a random sample of 100  issue reports.
Listing \ref{list:pilot} shows the process for the pilot analysis.
\begin{figure}[t]
	\centering
	\begin{lstlisting}[caption={Steps to perform the pilot analysis.}, label={list:pilot}]
Inputs = 100 issue reports, list of issue types from the selected taxonomies, list of issue types for the baseline taxonomy (initially empty).

For each issue report:
Manually analyze the contents of the issue report.
If the issue report matches existing issue type:
Add the issue type(s) to the baseline taxonomy.

Output = the baseline taxonomy.
	\end{lstlisting}
\end{figure}
Once the pilot analysis was complete, we were able to define the baseline taxonomy using the selected taxonomies.
For the GUI characteristic of the \canvas, most  issue types from the taxonomy provided by Lelli et al.~\cite{lelli2015classifying} were adopted into the baseline taxonomy.
For the web characteristic of the \canvas, one of the  issue types from the taxonomy provided by Marchetto et al.~\cite{marchetto2009empirical} was adopted into the baseline taxonomy.
The baseline taxonomy can be seen in Table~\ref{tab:baselineTaxonomy}.
\begin{table}[t]
	\centering
	\caption{The baseline taxonomy.}
	\begin{tabular*}{\linewidth}{l @{\extracolsep{\fill}} l}
		\toprule
		\textbf{Characteristic}     & \textbf{Type}          \\
		\midrule
		User interface      & Data presentation              \\
		& GUI structure and aesthetics   \\  
		\midrule               
		User interaction    & Action                         \\
		& Behaviour                      \\ 
		\midrule
		Web architecture    & Browser incompatibility        \\                          
		\bottomrule
	\end{tabular*}
	\label{tab:baselineTaxonomy}
\end{table}

\subsection{Constructing the taxonomy of testable \canvas  issues}\label{sec:construction}
To construct our taxonomy of testable \canvas  issues, we manually classified a sample of the  issue reports that we collected.
We randomly selected a statistically representative sample of 332 of our collected  issue reports with a confidence level of $95\%$ and a confidence interval of $5\%$.
We used stratified sampling to reduce bias~\cite{de2015investigating} towards projects with a higher number of  issue reports in our dataset. 
To manually classify the  issue reports, we performed a manual labelling process over two stages.
First, we performed multi-label classification to determine which  issue reports contained  issue types already defined in the baseline taxonomy.
Then, we performed card sorting primarily to elicit and define new issue types.
When assigning labels to issue reports, we considered not only the title and description provided by the reporter, but also discussion by developers and users in the comments of the issue reports.

\subsubsection{Independent multi-label classification} \label{sec:multilabel}
In the first stage of manual classification, the first and second authors each independently performed a multi-label classification of the sampled  issue reports.
Listing \ref{list:classify1} shows the process we followed to perform multi-label classification.
There were cases where no  issue types were assigned to an  issue report, because the baseline taxonomy was not fully descriptive of testable \canvas  issue types.
\begin{figure}[t]
	\begin{lstlisting}[caption={Steps to perform multi-label classification.}, label={list:classify1}]
Inputs = 332 issue reports, list of issue types from the baseline taxonomy.

For each issue report:
Independently analyze the contents of the issue report.
If the issue report matches existing issue type:
Label the issue report with relevant issue type(s).   

Output = 332 issue reports (partially classified).
	\end{lstlisting}
\end{figure}
To  measure inter-rater reliability (IRR) when performing multi-label classification of the issue reports, we calculated IRR on a per label basis, similar to prior work~\cite{viggiato2021causes}. 
We measured IRR for each label individually using Cohen’s kappa ($\kappa$)~\cite{cohen1960coefficient}, which we calculated as follows:

\begin{align*}
	\kappa=\frac{p_o-p_e}{1-p_e} 
	&&
	p_o=\frac{\#\text{agreements}}{N}
	&&
	p_e=\frac{1}{N^2}\sum_k{n_{k1}n_{k2}}
\end{align*}

Where $p_o$ is the observed percentage of agreement, $p_e$ is the expected percentage of agreement based on how often raters assigned the categories, $N$ is the total number of samples, and $n$ is the number of samples assigned category $k$ by each rater 1 and 2. For our multi-label classification we are calculating Cohen’s kappa on a per-label basis, and so the set of categories considered when calculating Cohen’s kappa is simply $\{\text{yes}, \text{no}\}$, and we repeat the calculation for each label. The subtables contained in Table~\ref{tab:cm1} each show the confusion matrix and kappa value for each label after the first step of our classification process.

After completing the first stage of manual classification, the first and second authors agreed that the \textit{GUI structure and aesthetics} and \textit{Data presentation} labels were not mutually exclusive, as these labels often had to be grouped together as if they were the same label.
This is because the \canvas can act as both a GUI and a graphics container for web applications.
So, to improve our classification, we merged the labels from the \textit{User interface} characteristic of the baseline taxonomy, and for clarity we renamed this characteristic to \textit{Visual} in our taxonomy.
After merging, we calculated Cohen’s kappa for the \textit{Visual} label to be $0.537$, an increase over either of the original labels.
Then, the sixth author independently performed conflict resolution for the cases where only one of the first two authors selected a specific issue type for a given issue report.
The low kappa values measured in our first stage of classification indicated that additional steps of taxonomy construction were necessary, and not every  issue report had been labelled yet, leading to the second stage of our manual classification.

\begin{table}[t]
	\caption{Confusion matrices for the first stage of manual classification with 332 open-source GitHub issue reports.}
	\renewcommand{\arraystretch}{1.2}
	\begin{subtable}[t]{0.45\columnwidth}
		\centering
		\begin{tabular}{ l l r r }
			& & \multicolumn{2}{c}{Rater 2} \\
			& & Yes & No \\ \cline{3-4}
			\multirow{2}{*}{Rater 1} & \multicolumn{1}{l|}{Yes} &  \multicolumn{1}{r|}{48} & \multicolumn{1}{r|}{64} \\ \cline{3-4}
			& \multicolumn{1}{l|}{No} & \multicolumn{1}{r|}{21} & \multicolumn{1}{r|}{199} \\
			\cline{3-4}
		\end{tabular}
		\vspace{4pt}
		\caption{\textit{Data Presentation} ($\kappa$~=~0.368)}
		\label{tab:cm1_datapresentation}
	\end{subtable}
	\hfill
	\begin{subtable}[t]{0.45\columnwidth}
		\centering
		\begin{tabular}{ l l r r }
			& & \multicolumn{2}{c}{Rater 2} \\
			& & Yes & No \\ \cline{3-4}
			\multirow{2}{*}{Rater 1} & \multicolumn{1}{l|}{Yes} &  \multicolumn{1}{r|}{24} & \multicolumn{1}{r|}{26} \\ \cline{3-4}
			& \multicolumn{1}{l|}{No} & \multicolumn{1}{r|}{45} & \multicolumn{1}{r|}{237} \\
			\cline{3-4}
		\end{tabular}
		\vspace{4pt}
		\caption{\textit{GUI Structure and Aesthetics}  ($\kappa$~=~0.277)}
		\label{tab:cm1_guistructure}
	\end{subtable}
	\hfill
	\newline
	\vspace*{6pt}
	\newline
	\begin{subtable}[t]{0.45\columnwidth}
		\centering
		\begin{tabular}{ l l r r }
			& & \multicolumn{2}{c}{Rater 2} \\
			& & Yes & No \\ \cline{3-4}
			\multirow{2}{*}{Rater 1} & \multicolumn{1}{l|}{Yes} &  \multicolumn{1}{r|}{11} & \multicolumn{1}{r|}{12} \\ \cline{3-4}
			& \multicolumn{1}{l|}{No} & \multicolumn{1}{r|}{24} & \multicolumn{1}{r|}{285} \\
			\cline{3-4}
		\end{tabular}
		\vspace{4pt}
		\caption{\textit{Action}  ($\kappa$~=~0.323)}
		\label{tab:cm1_action}
	\end{subtable}
	\hfill
	\begin{subtable}[t]{0.45\columnwidth}
		\centering
		\begin{tabular}{ l l r r }
			& & \multicolumn{2}{c}{Rater 2} \\
			& & Yes & No \\ \cline{3-4}
			\multirow{2}{*}{Rater 1} & \multicolumn{1}{l|}{Yes} &  \multicolumn{1}{r|}{11} & \multicolumn{1}{r|}{6} \\ \cline{3-4}
			& \multicolumn{1}{l|}{No} & \multicolumn{1}{r|}{16} & \multicolumn{1}{r|}{299} \\
			\cline{3-4}
		\end{tabular}
		\vspace{4pt}
		\caption{\textit{Behaviour}  ($\kappa$~=~0.466)}
		\label{tab:cm1_behaviour}
	\end{subtable}
	\hfill
	\newline
	\vspace*{6pt}
	\newline
	\begin{subtable}[t]{0.45\columnwidth}
		\centering
		\begin{tabular}{ l l r r }
			& & \multicolumn{2}{c}{Rater 2} \\
			& & Yes & No \\ \cline{3-4}
			\multirow{2}{*}{Rater 1} & \multicolumn{1}{l|}{Yes} &  \multicolumn{1}{r|}{30} & \multicolumn{1}{r|}{11} \\ \cline{3-4}
			& \multicolumn{1}{l|}{No} & \multicolumn{1}{r|}{10} & \multicolumn{1}{r|}{281} \\
			\cline{3-4}
		\end{tabular}
		\vspace{4pt}
		\caption{\textit{Browser incompatibility}  ($\kappa$~=~0.705)}
		\label{tab:cm1_browserincompatibility}
	\end{subtable}
	\hfill
	\begin{subtable}[t]{0.45\columnwidth}
		\centering
		\begin{tabular}{ l l r r }
			& & \multicolumn{2}{c}{Rater 2} \\
			& & Yes & No \\ \cline{3-4}
			\multirow{2}{*}{Rater 1} & \multicolumn{1}{l|}{Yes} &  \multicolumn{1}{r|}{102} & \multicolumn{1}{r|}{57} \\ \cline{3-4}
			& \multicolumn{1}{l|}{No} & \multicolumn{1}{r|}{19} & \multicolumn{1}{r|}{154} \\
			\cline{3-4}
		\end{tabular}
		\vspace{4pt}
		\caption{\textit{Visual} ($\kappa$~=~0.537)}
		\label{tab:cm1_visual}
	\end{subtable}
	\hfill
	\label{tab:cm1}
\end{table}

\subsubsection{Card sorting} \label{sec:cardsorting}
In the second stage of manual classification, the first and second authors performed card sorting together to develop a more comprehensive and detailed taxonomy of testable \canvas  issues.
Card sorting is a process of assigning each subject (issue report) to a single label, and is a widely-used approach that can help develop useful classifications~\cite{nurmuliani2004using}.
Listing \ref{list:classify3} shows the steps we took to perform card sorting.
\begin{figure}[!t]
	\begin{lstlisting}[caption={Steps to perform card sorting.}, label={list:classify3}]
Inputs = 332 issue reports (partially classified), list of issue types from our taxonomy. 

For each issue report:
Discuss the contents of the issue report.
If the issue report matches existing issue type:
Label the issue report with the most relevant issue type.
Else if the issue report does not contain a testable issue:
Discard the issue report.
Else:
Add a new issue type to the taxonomy.
Restart card sorting with updated taxonomy.

Output = the taxonomy of testable <canvas> issues.
	\end{lstlisting}
\end{figure}
Through card sorting, we were able to define new  issue types for our taxonomy and develop a more detailed classification for \textit{Visual}  issues.
After the first stage of manual classification, there were  issue reports that had been assigned more than one label because multiple  symptoms could be present in a single  issue report. During card sorting (the second stage of classification) we came to an agreement together to determine what the main  issue symptom (type) was, allowing us to assign a single label for such  issue reports.
9\% of the  issue reports did not contain a testable issue, and were discarded from the sample.

\subsection{Evaluating our taxonomy of testable \canvas  issues} \label{sec:evaluation}
After completing the construction of our taxonomy, we evaluated our final taxonomy using established quality criteria for taxonomies~\cite{usman2017taxonomies,ralph2018toward}.

\paragraph*{\textit{Utility demonstration}}
The utility of a taxonomy is demonstrated by classifying subject matter examples~\cite{usman2017taxonomies,vsmite2014empirically}, such as issue reports. To evaluate the utility of our taxonomy, the first and second author each independently performed a multi-label classification of a random sample of 50 issue reports from a proprietary \canvas application. The proprietary application is a \canvas game for children’s education that is played by more than 100 million children across the world. The game is representative of \canvas applications as the entire game is contained within the \canvas. We did not require any new labels to classify these issue reports, and measured considerably higher Cohen’s kappa values for all labels when compared to the first stage of our taxonomy construction. We calculated Cohen's kappa for our validation exercise the same way as described in Section~\ref{sec:multilabel}. Confusion matrices and kappa values for the validation exercise are shown in Table~\ref{tab:cm2}. Other than the \textit{Rendering} label, we measured the lowest kappa for a label to be 0.790 and the highest to be 1. For the \textit{Rendering} label, we measured a kappa value of 0 despite there only being a single disagreement as seen in Table~\ref{tab:cm2_rendering}. For the proprietary application, \textit{Rendering} issues are not a concern and are rarely reported, meaning that a single instance with disagreement led to a low kappa value. Our validation results show that our taxonomy is useful in describing testable \canvas issues found in real, industrial \canvas applications, and therefore meets the utility criterion.

\paragraph*{\textit{Orthogonality demonstration/Reflects (dis)similarities between instances}}
The orthogonality of a taxonomy, i.e., that the labels in the taxonomy do not overlap with each other, is implied by the design of the taxonomy~\cite{usman2017taxonomies,vsmite2014empirically}. Our results in Section~\ref{sec:resolutionresults} indicate that there is a difference in resolution times between our  testable issue types, which supports that there are differences between the  issue types in our taxonomy. However, during the evaluation of the utility of our taxonomy,  we observed that the industry issue reports sometimes required multiple labels. The need for multiple labels suggests that symptoms can occur together, and that the  issue types in our taxonomy are not completely mutually exclusive. At the same time, the symptoms do not always occur together; indicating that the  issue types in our taxonomy are dissimilar enough to warrant not merging them. As the goal of our taxonomy is to identify directions for research on \canvas testing, we do not consider it problematic that the taxonomy is not completely orthogonal. 

\paragraph*{\textit{Benchmarking}}
Benchmarking involves comparing the constructed taxonomy with similar classification schemes~\cite{usman2017taxonomies,vsmite2014empirically}. As described in Section~\ref{sec:pilot}, we conducted a pilot analysis to determine the relevance of labels in similar taxonomies (for GUI~\cite{lelli2015classifying} and web~\cite{marchetto2009empirical}  issues) when constructing our taxonomy of testable \canvas  issues, meaning our methodology for taxonomy construction contains an implicit benchmarking exercise with the two aforementioned taxonomies. Additionally, in Section~\ref{sec:discussion} of our paper, we provide detailed discussion of the (dis)similarities between our final taxonomy and the GUI~\cite{lelli2015classifying} and web~\cite{marchetto2009empirical}  issue taxonomies, including discussion of new  testable issue types in our taxonomy that are specific to the \canvas.

\paragraph*{\textit{Fit-for-purpose}}
To determine if our taxonomy is fit-for-purpose, we consider the degree to which our taxonomy is effective at the specific purpose it was designed for~\cite{ralph2018toward}. As defined in Section 1, our guiding question in constructing our taxonomy was \textit{``What types of  testable issues do developers encounter when creating web applications with the HTML5 \canvas?''}, and our underlying motivation is to direct future research on testing the \canvas. As discussed in Section~\ref{sec:summary}, our taxonomy answers our research question by providing a detailed classification of  testable issue types reported in popular open-source projects that utilize the \canvas, including  issue types defined in existing GUI~\cite{lelli2015classifying} and web~\cite{marchetto2009empirical}  issue taxonomies, as well as new  issue types specific to the \canvas. Furthermore, in Section~\ref{sec:futuredirections} we identified several future lines of research based on our taxonomy of testable \canvas issues. Based on these characteristics, we conclude that our taxonomy is fit-for-purpose.

\begin{table}[t]
	\caption{Confusion matrices for the validation exercise with 50 industry issue reports.}
	\renewcommand{\arraystretch}{1.2}
	\begin{subtable}[t]{0.45\columnwidth}
		\centering
		\begin{tabular}{ l l r r }
			& & \multicolumn{2}{c}{Rater 2} \\
			& & Yes & No \\ \cline{3-4}
			\multirow{2}{*}{Rater 1} & \multicolumn{1}{l|}{Yes} &  \multicolumn{1}{r|}{9} & \multicolumn{1}{r|}{2} \\ \cline{3-4}
			& \multicolumn{1}{l|}{No} & \multicolumn{1}{r|}{0} & \multicolumn{1}{r|}{39} \\
			\cline{3-4}
		\end{tabular}
		\vspace{4pt}
		\caption{\textit{Layout}  ($\kappa$~=~0.875)}
		\label{tab:cm2_layout}
	\end{subtable}
	\hfill
	\begin{subtable}[t]{0.45\columnwidth}
		\centering
		\begin{tabular}{ l l r r }
			& & \multicolumn{2}{c}{Rater 2} \\
			& & Yes & No \\ \cline{3-4}
			\multirow{2}{*}{Rater 1} & \multicolumn{1}{l|}{Yes} &  \multicolumn{1}{r|}{15} & \multicolumn{1}{r|}{4} \\ \cline{3-4}
			& \multicolumn{1}{l|}{No} & \multicolumn{1}{r|}{0} & \multicolumn{1}{r|}{31} \\
			\cline{3-4}
		\end{tabular}
		\vspace{4pt}
		\caption{\textit{State}  ($\kappa$~=~0.823)}
		\label{tab:cm2_state}
	\end{subtable}
	\hfill
	\newline
	\vspace*{6pt}
	\newline
	\begin{subtable}[t]{0.45\columnwidth}
		\centering
		\begin{tabular}{ l l r r }
			& & \multicolumn{2}{c}{Rater 2} \\
			& & Yes & No \\ \cline{3-4}
			\multirow{2}{*}{Rater 1} & \multicolumn{1}{l|}{Yes} &  \multicolumn{1}{r|}{1} & \multicolumn{1}{r|}{0} \\ \cline{3-4}
			& \multicolumn{1}{l|}{No} & \multicolumn{1}{r|}{0} & \multicolumn{1}{r|}{49} \\
			\cline{3-4}
		\end{tabular}
		\vspace{4pt}
		\caption{\textit{Appearance}  ($\kappa$~=~1)}
		\label{tab:cm2_appearance}
	\end{subtable}
	\hfill
	\begin{subtable}[t]{0.45\columnwidth}
		\centering
		\begin{tabular}{ l l r r }
			& & \multicolumn{2}{c}{Rater 2} \\
			& & Yes & No \\ \cline{3-4}
			\multirow{2}{*}{Rater 1} & \multicolumn{1}{l|}{Yes} &  \multicolumn{1}{r|}{0} & \multicolumn{1}{r|}{0} \\ \cline{3-4}
			& \multicolumn{1}{l|}{No} & \multicolumn{1}{r|}{1} & \multicolumn{1}{r|}{49} \\
			\cline{3-4}
		\end{tabular}
		\vspace{4pt}
		\caption{\textit{Rendering}  ($\kappa$~=~0)}
		\label{tab:cm2_rendering}
	\end{subtable}
	\hfill
	\newline
	\vspace*{6pt}
	\newline
	\begin{subtable}[t]{0.45\columnwidth}
		\centering
		\begin{tabular}{ l l r r }
			& & \multicolumn{2}{c}{Rater 2} \\
			& & Yes & No \\ \cline{3-4}
			\multirow{2}{*}{Rater 1} & \multicolumn{1}{l|}{Yes} &  \multicolumn{1}{r|}{9} & \multicolumn{1}{r|}{1} \\ \cline{3-4}
			& \multicolumn{1}{l|}{No} & \multicolumn{1}{r|}{1} & \multicolumn{1}{r|}{39} \\
			\cline{3-4}
		\end{tabular}
		\vspace{4pt}
		\caption{\textit{Browser runtime error} ($\kappa$~=~0.875)}
		\label{tab:cm2_runtimeerror}
	\end{subtable}
	\hfill
	\begin{subtable}[t]{0.45\columnwidth}
		\centering
		\begin{tabular}{ l l r r }
			& & \multicolumn{2}{c}{Rater 2} \\
			& & Yes & No \\ \cline{3-4}
			\multirow{2}{*}{Rater 1} & \multicolumn{1}{l|}{Yes} &  \multicolumn{1}{r|}{1} & \multicolumn{1}{r|}{0} \\ \cline{3-4}
			& \multicolumn{1}{l|}{No} & \multicolumn{1}{r|}{0} & \multicolumn{1}{r|}{49} \\
			\cline{3-4}
		\end{tabular}
		\vspace{4pt}
		\caption{\textit{Different behaviour across browsers}  ($\kappa$~=~1)}
		\label{tab:cm2_differentbehaviour}
	\end{subtable}
	\hfill
	\newline
	\vspace*{6pt}
	\newline
	\begin{subtable}[t]{0.45\columnwidth}
		\centering
		\begin{tabular}{ l l r r }
			& & \multicolumn{2}{c}{Rater 2} \\
			& & Yes & No \\ \cline{3-4}
			\multirow{2}{*}{Rater 1} & \multicolumn{1}{l|}{Yes} &  \multicolumn{1}{r|}{9} & \multicolumn{1}{r|}{0} \\ \cline{3-4}
			& \multicolumn{1}{l|}{No} & \multicolumn{1}{r|}{0} & \multicolumn{1}{r|}{41} \\
			\cline{3-4}
		\end{tabular}
		\vspace{4pt}
		\caption{\textit{Action}  ($\kappa$~=~1)}
		\label{tab:cm2_action}
	\end{subtable}
	\hfill
	\begin{subtable}[t]{0.45\columnwidth}
		\centering
		\begin{tabular}{ l l r r }
			& & \multicolumn{2}{c}{Rater 2} \\
			& & Yes & No \\ \cline{3-4}
			\multirow{2}{*}{Rater 1} & \multicolumn{1}{l|}{Yes} &  \multicolumn{1}{r|}{4} & \multicolumn{1}{r|}{0} \\ \cline{3-4}
			& \multicolumn{1}{l|}{No} & \multicolumn{1}{r|}{1} & \multicolumn{1}{r|}{45} \\
			\cline{3-4}
		\end{tabular}
		\vspace{4pt}
		\caption{\textit{Behaviour}  ($\kappa$~=~0.878)}
		\label{tab:cm2_behaviour}
	\end{subtable}
	\hfill
	\newline
	\vspace*{6pt}
	\newline
	\begin{subtable}[t]{0.45\columnwidth}
		\centering
		
		\begin{tabular}{ l l r r }
			& & \multicolumn{2}{c}{Rater 2} \\
			& & Yes & No \\ \cline{3-4}
			\multirow{2}{*}{Rater 1} & \multicolumn{1}{l|}{Yes} &  \multicolumn{1}{r|}{2} & \multicolumn{1}{r|}{0} \\ \cline{3-4}
			& \multicolumn{1}{l|}{No} & \multicolumn{1}{r|}{1} & \multicolumn{1}{r|}{47} \\
			\cline{3-4}
		\end{tabular}
		\vspace{4pt}		
		\caption{\textit{Inefficient memory usage}  ($\kappa$~=~0.790)}
		\label{tab:cm2_inefficient memory usage}
	\end{subtable}
	\hfill
	\label{tab:cm2}
\end{table}

\subsection{Determining  issue resolution time} \label{sec:resolution}
In addition to constructing a taxonomy of testable \canvas  issues, we also quantified  issue resolution time for our set of classified open-source issue reports to better understand how developers deal with different types of testable \canvas issues. 
To calculate the issue resolution time, we calculated the difference between the issue open time and the final issue close time for each  issue report, similar to prior work~\cite{bogner2022type}. We calculated the median resolution time for each  issue characteristic in our taxonomy by taking the median of the resolution times for all  issue reports associated with each  issue characteristic.

\subsubsection{Collecting data for comparison} \label{sec:collectingcomparison}
To provide a baseline for comparison for  issue resolution times, we collected non-\canvas  issue reports from the projects in our studied population. We followed a similar process to how we identified \canvas  issue reports in the projects, except we collected only  issue reports that did not mention the \canvas.
	
\begin{table*}[!t]
	\centering
	\caption{Our taxonomy of testable \canvas  issues. Newly defined  issue types are shown in bold font.}
	\begin{tabular*}{\textwidth}{l @{\extracolsep{\fill}} l @{\extracolsep{\fill}} l}
		\toprule
		\textbf{Characteristic}   & \textbf{Type}                & \textbf{Description} \\
		\midrule
		Visual (35\%)           & Rendering (20\%)                              & Objects appear blurry, distorted, or contain artifacts. \\
		& Layout (12\%)                                 & Objects have incorrect positioning, layering, or size. \\     
		& State (2\%)                                   & Objects displayed in the wrong state, e.g. visible vs. invisible. \\
		& Appearance (1\%)                              & Objects have incorrect aesthetics, e.g. wrong colour. \\                      
		\midrule
		User interaction (14\%) & Action (6\%)                                  & Incorrect result from single DOM event on the \canvas. \\
		& Behaviour (8\%)                               & Incorrect result from combination of DOM events on the \canvas. \\
		\midrule
		Web architecture (17\%) & Different behaviour across browsers (14\%)                & Incorrect \canvas functionality on specific browsers. \\
		& \textbf{Cross-origin resource sharing} (3\%)  & Incorrect usage of CORS policies prevents the loading of resources.\\                 
		\midrule
		Performance (5\%)      & \textbf{Inefficient memory usage} (5\%)    & Memory leaks from inefficient usage of the Canvas or WebGL APIs. \\                     
		\midrule
		Integration (29\%)      & \textbf{Saving \canvas data} (2\%)            & Saving the \canvas bitmap to a file provides an unexpected result. \\
		& \textbf{Browser runtime error} (27\%)         & Error outside of the \canvas prevents \canvas functionality. \\                                        
		\bottomrule
	\end{tabular*}
	\label{tab:taxonomy}
\end{table*}

\section{Our taxonomy of testable canvas  issues} \label{sec:ourtaxonomy}
In this section, we provide a detailed description of each of the  types in our taxonomy of testable \canvas  issues.
Table~\ref{tab:taxonomy} shows our taxonomy of testable \canvas  issues, including the frequency of each  issue type in the studied  issue reports.

\subsection{Visual  issues}
\textit{Visual}  issues are problems in the presentation of objects on the \canvas bitmap.
We discovered four types of \textit{Visual}  issues: \textit{Rendering}, \textit{Layout}, \textit{State}, and \textit{Appearance}.

\begin{figure*}[t]
	\centering
	\begin{subfigure}[t]{0.2\textwidth}
		\centering
		\includegraphics[width=\textwidth]{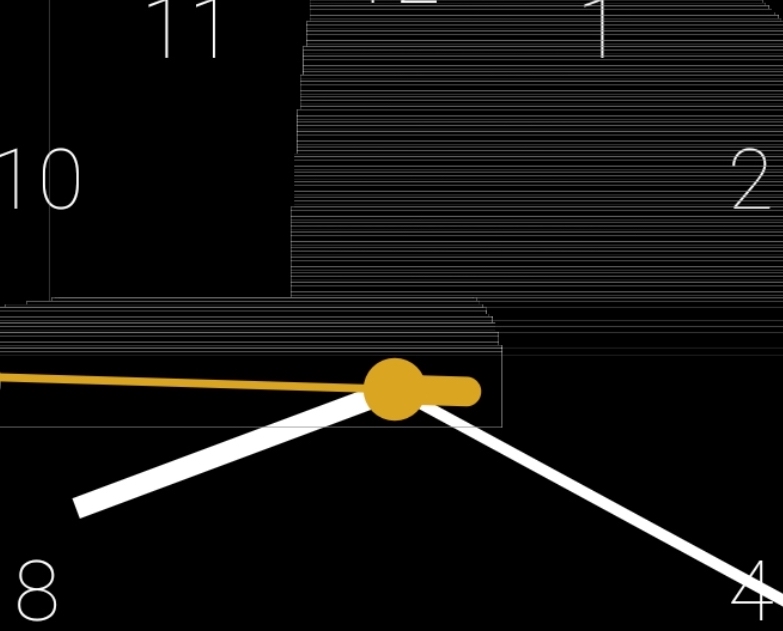}
		\caption{\textit{Rendering}  issue (clock contains artifacts)~\cite{issuereport1}.}
		\label{fig:rendering}
	\end{subfigure}
	\hfill
	\begin{subfigure}[t]{0.25\textwidth}
		\centering
		\includegraphics[width=\textwidth]{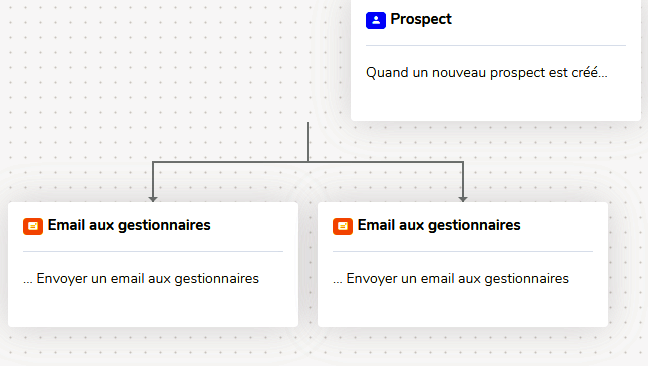}
		\caption{\textit{Layout}  issue (object has incorrect position)~\cite{issuereport10}.}
		\label{fig:layout}
	\end{subfigure}
	\hfill
	\begin{subfigure}[t]{0.25\textwidth}
		\centering
		\includegraphics[width=\textwidth]{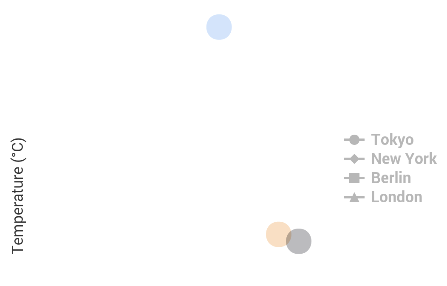}
		\caption{\textit{State}  issue (circles should be hidden)~\cite{issuereport15}.}
		\label{fig:state}
	\end{subfigure}
	\hfill
	\begin{subfigure}[t]{0.2\textwidth}
		\centering
		\includegraphics[width=\textwidth]{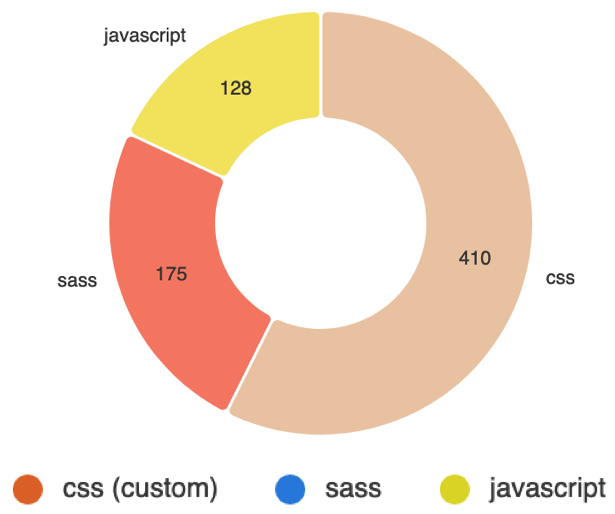}
		\caption{\textit{Appearance}  issue (wrong colours)~\cite{issuereport18}.}
		\label{fig:appearance}
	\end{subfigure}
	\caption{Sample instances of \textit{Visual}  issues reported in the open source projects.}
	\label{fig:bug_samples}
\end{figure*}

\textit{Rendering}  issues are often related to the incorrect scaling of objects or images on the \canvas, with example instances such as distortion or blurriness being common.
Some instances of \textit{Rendering}  issue reports described unexpected visual results due to more specific logical errors in using the Canvas API (or WebGL API) to display specific shapes, such as a waveform~\cite{issuereport8}.
Figure~\ref{fig:rendering} shows an example instance of a \textit{Rendering}  issue~\cite{issuereport1}.
We noticed that some \textit{Rendering}  issue reports discussed how the incorrect scaling occurred due to a failure in accounting for differences in device resolutions~\cite{issuereport2, issuereport3, issuereport4, issuereport5}, or changes in \canvas dimensions~\cite{issuereport6, issuereport7}.
We also found an instance of a \textit{Rendering}  issue where artifacts appeared on the screen only when using WebGL on devices with an outdated graphics driver~\cite{issuereport9}.

\textit{Layout}  issues are related to incorrect positioning and sizing of objects on the \canvas.
Figure \ref{fig:layout} shows an example instance of a \textit{Layout}  issue, with misaligned objects on the \canvas~\cite{issuereport10}.
\textit{Layout}  issues may be the incorrect coordinates (like $[x, y]$) for an object~\cite{issuereport11}, or the incorrect layering (like z-index) for an object~\cite{issuereport12} on the \canvas.
In one instance, a data visualization was not fully viewable because some objects had been assigned positions outside of the \canvas viewport~\cite{issuereport13}.
In another instance, a \canvas animation was not fully viewable due to the \canvas dimensions being too small~\cite{issuereport14}.

\textit{State}  issues are those in which an object (or text) is displayed in a different state than it should be.
For example, an object that is visible on the \canvas should be invisible, or vice versa.
Figure \ref{fig:state} shows an example instance of a \textit{State}  issue, with circles appearing when they should not be visible~\cite{issuereport15}.
In one \textit{State}  issue report, game controls were not removed when changing views on the \canvas to a text box for chat~\cite{issuereport16}.
In another more complicated case, multiple HTML \canvas elements were used to render different objects, and specific HTML \canvas elements were not being hidden when they should have been \cite{issuereport17}.

\textit{Appearance}  issues are related to the incorrect aesthetics of objects on the \canvas, such as incorrect colour, transparency, or font.
Figure \ref{fig:appearance} shows an example instance of an \textit{Appearance}  issue, with the legend of the graph having incorrect colours~\cite{issuereport18}.
We found some instances of \textit{Appearance}  issues where the incorrect font was used when rendering text on the \canvas~\cite{issuereport19, issuereport20}.

\subsection{User interaction  issues}
\textit{User interaction}  issues are related to DOM events that are triggered on the HTML \canvas element.
We discovered two types of \textit{User interaction} \canvas  issues.
\textit{Action}  issues occur when a single action fails to execute correctly.
Meanwhile, \textit{Behaviour}  issues are those which consist of multiple actions that together provide an incorrect result.
Some example instances of \textit{Action}  issue reports we encountered include: click-dragging the cursor to zoom does not work when the cursor moves outside of the bounds of the HTML \canvas element~\cite{issuereport21}, and a tooltip is not showing when mouse hovering over a graph rendered to the \canvas~\cite{issuereport22}.
Other instances of \textit{Action}  issue reports were similar in that a specific action was failing~\cite{issuereport23, issuereport24, issuereport25},\label{key} which prevented user interaction functionality in web applications built with the \canvas.
For \textit{Behaviour}  issues, some example instances we encountered include: drawing functionality does not work correctly while simultaneously scrolling within the \canvas~\cite{issuereport26}, and dragging an object on the \canvas does not function as expected after scrolling within the \canvas~\cite{issuereport27}.
In general, \textit{Behaviour}  issues involve two or more actions that are performed simultaneously or in succession, which ultimately provide incorrect functionality with the \canvas. 
Although \textit{User interaction}  issues are related to DOM events, they were often described in  issue reports by comparing the expected and buggy visual results.

\subsection{Web architecture  issues}
We discovered two types of \textit{Web architecture} \canvas  issues: \textit{Different behaviour across browsers}, and \textit{Cross-origin resource sharing}.

\textit{Different behaviour across browsers}  issues occur when a \canvas application works on some browsers but not others.
Many such  issues were caused by the use of an outdated browser, such as IE8~\cite{issuereport28}, or an outdated version of a modern browser, such as Firefox 51.0 when the latest stable release was Firefox 63.0~\cite{issuereport29}.
However, we also found instances of \textit{Different behaviour across browsers}  issues where \canvas functionality was incorrect for a latest stable release version of a modern browser.
For example, we found some instances where a \canvas application did not function correctly on Safari, despite working on Firefox and Chrome~\cite{issuereport30, issuereport31}.
In another case, a specific browser  issue in Chrome prevented a \canvas application from functioning~\cite{issuereport32}.

\textit{Cross-origin resource sharing} (CORS)  issues are related to the incorrect use of CORS policies with \canvas.
CORS policies provide a secure method for loading \canvas resources (such as images or WebGL textures) from foreign domains, but if they are not used correctly, the resources will either not be loaded~\cite{mozillacors}, or prevent the \canvas data from being saved to a file~\cite{mozillacorsimage}.
In one instance, an AWS bucket was not correctly configured with CORS policies, which prevented the loading of those resources to a \canvas application~\cite{issuereport33}.
In some instances, CORS headers were not used in the HTTP requests to foreign domains, preventing resources from being loaded~\cite{issuereport34, issuereport35}.

\subsection{Performance issues}
\textit{Performance} issues can occur with the \canvas due to \textit{Inefficient memory usage}.
issue reports which described \textit{Inefficient memory usage}  issues commonly discussed how the outcome of such bugs was a frozen or crashed web application.
Some example instances of \textit{Inefficient memory usage}  issues include \canvas applications crashing due to: rendering occurring too often~\cite{issuereport36}, and new objects being rendered to the \canvas without old (hidden/invisible) objects being removed from memory~\cite{issuereport37}.

\subsection{Integration issues}
When connecting the \canvas with other parts of a web application, \textit{Integration}  issues can occur.
We discovered two types of \textit{Integration}  issues: \textit{Saving \canvas data} and \textit{Browser runtime error}.

\textit{Saving \canvas data}  issues can occur when saving the \canvas bitmap to a file using custom JavaScript methods or JavaScript methods built into the browser.
One instance described how saving the \canvas bitmap using a built-in browser method produced a file with a black background that was not present in the browser~\cite{issuereport38}.
This indicates that built-in methods such as \texttt{getImageData}, \texttt{toBlob}, or \texttt{toDataURL} may not produce the same result as what is seen on the \canvas.
Other instances of \textit{Saving \canvas data} bugs were due to errors in custom methods for saving the \canvas bitmap, for example we found an instance where the saved data was unexpectedly compressed~\cite{issuereport39}.

We discovered  issue reports that discussed \canvas functionality despite the  issue being located in a different part of the web application, and classified such  issue reports as \textit{Browser runtime error}.
While these  issue reports did mention the \canvas, these  issues were not actually directly related to \canvas problems.
Example instances of \textit{Browser runtime error}  issues include: referencing an HTML \canvas element that had not been created~\cite{issuereport40}, and using incorrect syntax for a JavaScript library method that added objects to a data visualization on the \canvas~\cite{issuereport41}.
It appears that most \textit{Browser runtime error}  issues are reported by developers who are not experienced with web development or the \canvas.

\subsection{Results for issue resolution time} \label{sec:resolutionresults}
As described in Section~\ref{sec:resolution}, we quantified the median  issue resolution times for each  issue characteristic in our taxonomy to better understand how developers deal with different types of testable \canvas  issues. Table~\ref{tab:resolution_times} shows the median  issue resolution times per  issue characteristic in our taxonomy. Our results indicate that developers of open source projects that utilize the \canvas may, for example, prioritize  issues reports belonging to \textit{Visual} bug types over  issue reports belonging to \textit{Web Architecture} bug types.

\subsubsection{Resolution times for non-\canvas  issue reports}
As described in Section~\ref{sec:collectingcomparison}, we also collected non-\canvas  issue reports to compare  issue resolution times. For the non-\canvas  issue reports in the 123 studied projects, we found that the median resolution time was 6 days. Comparing this result with our results in Table~\ref{tab:resolution_times}, we can conclude that developers may either prioritize non-\canvas  issue reports over most \canvas  issue reports, or that \canvas  issue reports are harder to resolve. However, \textit{Integration}  issues (from our taxonomy) are usually resolved more quickly than non-\canvas  issue reports.

\begin{table}[!t]
	\centering
	\caption{Median  issue resolution time per characteristic in our taxonomy.}
	\begin{tabular*}{\linewidth}{l @{\extracolsep{\fill}} r}
		\toprule
		\textbf{Characteristic in our taxonomy} & \textbf{Median time to resolve (days)} \\
		\midrule
		Integration & 4 \\
		Visual & 8 \\
		User Interaction & 13 \\
		Performance & 32 \\
		Web Architecture & 48 \\
		\bottomrule
	\end{tabular*}
	\label{tab:resolution_times}
\end{table}
	
\section{Discussion} \label{sec:discussion}
\begin{figure}[!t]
	\centering
	\includegraphics[width=\linewidth]{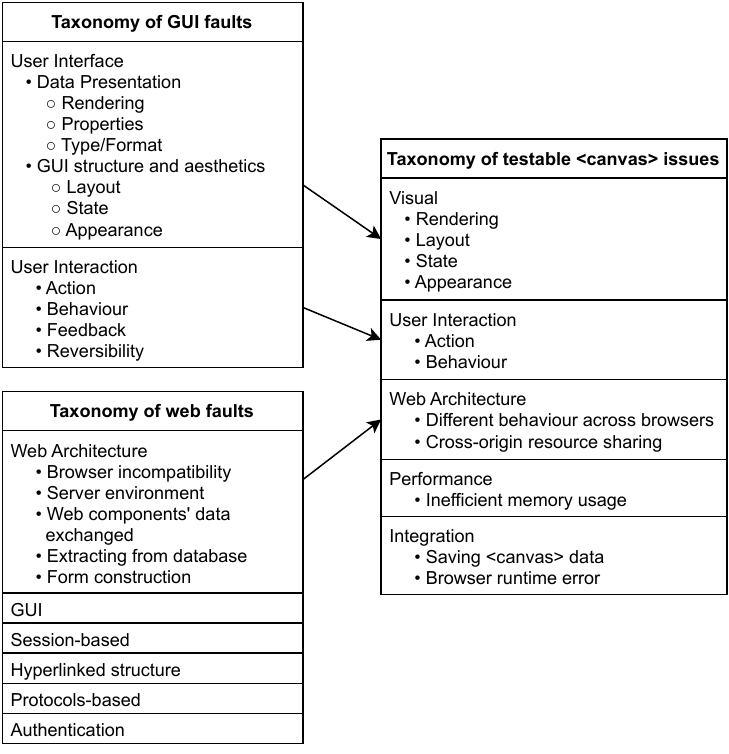}
	\caption{Taxonomy of GUI  faults~\cite{lelli2015classifying} and taxonomy of web  faults~\cite{marchetto2009empirical} mapped to our taxonomy.}
	\label{fig:discussion}
\end{figure}

Prior to performing classification, we hypothesized that testable \canvas  issues are similar to the  issues found in GUIs and web applications.
While we did discover some overlap, we found that some GUI  issue types and many web  issue types are not relevant to \canvas applications.
We also discovered some new types of  issues that are specific to the \canvas.
In this section, we discuss the differences between our taxonomy of testable \canvas  issues and the taxonomies that contributed to the baseline taxonomy (that we selected in Section \ref{sec:baseline}).
A mapping of the selected taxonomies to our taxonomy of testable \canvas  issues can be seen in Figure~\ref{fig:discussion}.

\subsection{Comparing GUI  issues and \canvas  issues}
The key difference between GUI  issues and \canvas  issues is that it is difficult to distinguish \textit{GUI structure and aesthetics} and \textit{Data presentation} for the \canvas.
Additionally, while several of the specific \textit{User interface} issue types provided in the taxonomy of GUI  faults~\cite{lelli2015classifying} were descriptive of \textit{Visual} issues on the \canvas, we did not use \textit{Type/format} or \textit{Properties} in our classification.
The absence of such  issue types in our taxonomy may be explained by the fact that the \canvas renders a bitmap rather than a structured markup language (e.g. HTML, XML) that contains separate properties for the display of raw numerical and textual data.
With regards to \textit{User interaction}  issue types, we did not find any \textit{Feedback} or \textit{Reversibility}  issues in \canvas  issue reports.
Together, these comparisons indicate that the \canvas is like a specific kind of GUI that deals heavily with graphics rendering.
However, our taxonomy of testable \canvas  issues contains  issue types that are not GUI-related, meaning \canvas testing has additional complexities that are not seen in desktop GUIs. 

\subsection{Comparing web  issues and \canvas  issues}
Our taxonomy of testable \canvas  issues contains very few  issue types seen in the taxonomy of web  faults~\cite{marchetto2009empirical}, indicating there is not a lot of overlap between \canvas issues and generic web application issues.
For the \textit{Web architecture} aspect of \canvas, the only obvious similarity between the taxonomy of web  faults~\cite{marchetto2009empirical} and our taxonomy of testable \canvas issues is the presence of the \textit{Different behaviour across browsers} type.
Given that \canvas is a client-side technology (i.e. it is implemented by browsers), the existence of \textit{Different behaviour across browsers} issues is not surprising.
In our taxonomy of testable \canvas  issues, the \textit{Cross-origin resource sharing} type is somewhat similar to the \textit{Server environment} type in the taxonomy of web issues~\cite{marchetto2009empirical}.
Knowing that \canvas is a client-side technology, one might not expect \textit{Server environment} issues to affect \canvas in web applications.
However, the specific challenges posed by CORS means that there are some \canvas  issues that could also be described by \textit{Server environment}. 
We did not use \textit{Web components' data exchanged}, \textit{Extracting from database}, or \textit{Form construction} types, meaning \canvas  issues do not range the full scope of generic web architecture  issues.
The smaller scope of web-related \canvas  issues compared to generic web application  issues is also indicated by the lack of many other types that were proposed in the taxonomy~\cite{marchetto2009empirical} from which we took initial \textit{Web architecture} types.

\subsection{Summary of findings} \label{sec:summary}
To complete the discussion of our constructed taxonomy of testable \canvas  issues, below we summarize our findings with reference to the research question defined in the introduction section of our paper: \textit{What types of  testable issues do developers encounter when creating web applications with the HTML5 \canvas?}

\begin{tcolorbox}
	\textit{Summary:} As our taxonomy demonstrates, developers encounter a wide variety of  testable issue types when creating web applications with the HTML5 \canvas. These  issue types span various characteristics of the \canvas, such as \textit{Visual}, \textit{Performance}, and \textit{Web Architecture}  issues. It is unlikely that there exists a silver bullet testing approach to cover this variety of testable \canvas  issues. Therefore, different \canvas testing approaches must be developed to test these different types of  issues.
\end{tcolorbox}

\section{Future research directions} \label{sec:futuredirections}

\subsection{Detecting \textit{Visual}  issues}

The most frequently reported \canvas  issues in the open source projects are \textit{Visual}  issues, as can be seen in Table~\ref{tab:taxonomy}.
\textit{Visual}  issues may be hard to diagnose, because the \canvas state is not exposed through the DOM.
The only existing research~\cite{bajammal2018web} that addresses testing the \canvas focuses on detecting what we have defined as \textit{State}  issues in our taxonomy.
Although there is extensive prior work regarding Visual GUI testing~\cite{alegroth2016maintenance,bajammal2020survey}, this testing paradigm is not synonymous with detecting \textit{Visual} issues~\cite{issa2012visual}. \textit{Visual} issues would interfere with Visual GUI testing methods that leverage the visual aspect of the application to drive test automation. Snapshot testing is a Visual GUI testing approach that could target \textit{Visual} issues on the \canvas, however snapshot testing has many inherent drawbacks that limit its utility as an automated testing strategy~\cite{bajammal2018web}, and future research is required to understand to what extent snapshot testing is useful in catching \textit{Visual} issues on the \canvas in web applications.
This emphasizes that research on \canvas testing is still in an early stage, and that there are many opportunities for future research on \canvas testing. 
Given the high frequency and notable variety of \textit{Visual} issues, new approaches for testing the \canvas must be developed.

\subsection{Detecting correct usage of the \canvas}

Using the \canvas can lead to new challenges in terms of performance.
While the \canvas opens the door to many interesting use-cases in web applications, it also may lead to \textit{Performance} issues that are specific to the \canvas.
As seen in Table~\ref{tab:taxonomy}, one of our newly defined types of \canvas  issues is \textit{Inefficient memory usage}, which can lead to memory leaks or high CPU usage.
While much of the graphics rendering with the \canvas is optimized under-the-hood, particularly when using WebGL, it is still up to developers to utilize the provided APIs in a performant manner.
The same \canvas animation might be produced through several different approaches, and some approaches outperform others.
In lieu of performance testing tools for the \canvas, developers may be able to utilize performance profiling tools built into a browser, such as Chrome DevTools~\cite{analyzeruntime}, to analyze the runtime performance of web applications built with the \canvas. Alternatively, an approach to profiling graphics performance such as the one detailed by Hoetzlein~\cite{hoetzlein2012graphics} could also prove useful. However, performance analysis is not an automated testing approach, and automated performance testing would be very difficult with performance profiling tools due to their computational overhead.
Therefore, future research should investigate the detection of \textit{Inefficient memory usage}  issues.

As shown in Table~\ref{tab:taxonomy}, one of our newly defined  issue types is \textit{Cross-origin resource sharing}.
To render resources (such as images or WebGL textures) from foreign domains to the \canvas, the foreign domains must be configured to allow CORS requests, and any HTTP request for these resources must contain a valid set of headers as defined in the CORS protocol~\cite{fetchspeccors}.
Failure to correctly implement either end of the communication will lead to these resources not being rendered to the \canvas.
Future research should develop testing approaches for the \canvas that detect incorrect usage of CORS policies with the \canvas.

Many  issues that appear to affect the \canvas are  issues that are \textit{Browser runtime errors}.
As shown in Table~\ref{tab:taxonomy}, many  issue reports in the open source projects discussed \canvas  issues that were not actually due to problems with \canvas, but instead runtime errors that consequently affected \canvas functionality.
When debugging web applications that use the \canvas, it would be useful to be able to automatically identify  issues that are \textit{Browser runtime errors}, so that developers do not waste time trying to locate \canvas  issues that originate in a different part of the web application.
Therefore, future research should aim to save developer time and effort by developing testing tools for the \canvas that help decide whether an  issue originated in or outside of the \canvas. 

\subsection{Detecting browser issues}

Some browsers may provide incorrect \canvas functionality.
As can be seen in Table~\ref{tab:taxonomy}, a fairly frequent type of \canvas  issue is \textit{Different behaviour across browsers}.
However, widely-used cross-browser testing frameworks rely on analyzing the DOM, meaning new forms of cross-browser testing are required for the \canvas.
Future research should investigate cross-browser testing methods for the \canvas.

Browser-provided JavaScript methods for saving \canvas data may not always provide the expected result.
As can be seen in Table~\ref{tab:taxonomy}, one of our newly defined types of \canvas  issue is \textit{Saving \canvas data}.
Although there are JavaScript methods built into browsers for saving the \canvas bitmap to a file (such as \textit{toDataURL} and \textit{getImageData}), we found one  issue report where a built-in browser method produced a different result than what was rendered on the web page.
This means that even if such methods usually work, we may not be able to fully rely on them to know what was rendered to a specific frame on the \canvas.
This finding has particular relevance for the development of \canvas testing tools that rely on visual analysis, which would require high confidence in the input data. 
Further research should seek to understand why browser-provided JavaScript methods for saving the \canvas bitmap can provide unexpected results.

\section{Threats to Validity} \label{sec:threats}

\paragraph*{\textit{Internal validity}}
Our project selection relied on the direct mention of \canvas-related keywords in each GitHub project's title, description, topic tags, or readme file.
While this provided a wide coverage of search results that had to be manually filtered for false positives, we may still have missed some \canvas projects, producing a possible bias in our results.
We mitigated this threat mostly by including a wide range of \canvas projects and using stratified sampling to reduce bias towards specific projects.
However, further studies may be required to confirm that our results generalize to all open source \canvas projects.

We used a custom set of keywords to identify  testable issue reports in the set of unlabelled GitHub issue reports, which poses a threat to internal validity.
While we may have missed some \canvas  issue reports, our set of keywords reflects terms often found in GitHub issue reports that report a  testable issue.
More research is required to better automatically identify testable issue reports in a set of GitHub issue reports.

Potential subjectivity in manually classifying  issue reports is a threat to internal validity.
While automatic classification could have reduced such subjectivity, manual classification was required to discover new testable issue types.
Additionally, we did not perform the second stage of our classification process independently, as this was necessary to perform card sorting and ensure we elicited all relevant labels for our taxonomy of \canvas  issues.
We addressed this threat first by selecting a baseline set of  issue types from empirically validated taxonomies that are related to known characteristics of the \canvas.
Then, the first two authors initially independently classified the  issue reports, which helps reduce bias and mitigate the potential subjectivity.
The first stage of our manual classification had fairly low Cohen’s kappa values for most labels, but this was not a concern as we completed further rounds of classification and measured high kappa values in our validation exercise. We then leveraged the results of our independent classification to perform card sorting with three authors. Having more than one author participate in card sorting also helped reduce potential subjectivity. Finally, we performed a validation exercise as described in Section~\ref{sec:evaluation} of our paper, in which we performed multi-label classification of 50 \canvas issue reports from a proprietary \canvas application, and measured high Cohen's kappa values for all labels except \textit{Rendering}.
Also, because we used a statistically representative random sample during classification, our results should be valid for the full set of $2,403$ collected issue reports.

A threat to internal validity is our choice of inter-rater reliability (IRR) measure in our independent multi-label classification and validation exercise. 
We measured IRR using Cohen’s kappa on a per-label basis for our multi-label classification (and validation exercise) because calculating Cohen’s kappa directly over all labels would be inconsistent with the labeling task and ignore cases of partial agreement~\cite{rosenberg2004augmenting}.
Although the use of Krippendorf’s alpha and variations of Cohen’s kappa (which account for coder bias) have been proposed to calculate agreement for multi-label classifications~\cite{artstein2008inter, rosenberg2004augmenting, passonneau2006measuring}, it is difficult to compare these agreement scores across different manual classification exercises~\cite{artstein2008inter}. For example, differences between choice of distance measure in Krippendorf’s alpha can lead to greater variability in agreement scores than coder bias alone in multi-label classifications~\cite{artstein2008inter}, and there is no consensus on which distance measure to use~\cite{artstein2008inter}. 
We have not provided any thresholds to interpret Cohen’s kappa, as there is a lack of consensus on how to interpret these IRR values~\cite{artstein2008inter}.

\paragraph*{\textit{External validity}}
Our findings are only valid for open-source \canvas projects on GitHub, and may not generalize to other \canvas projects.
Further studies should validate whether our results generalize to other \canvas projects.

To automatically identify \canvas  issue reports in our set of collected \canvas projects, we manually filtered out projects that utilized a canvas object that was not related to HTML5.
Examples of such projects include desktop \texttt{Node.js} applications that require \texttt{node-canvas}, or web applications that use \texttt{react-canvas} to render \texttt{React} components to \canvas instead of the DOM.
Further research would be required to validate that our findings generalize to projects that use non-HTML5 canvas objects.

\section{Conclusion} \label{sec:conclusion}
In this paper, we study the types of  testable issues that developers face when using the HTML5 \canvas in their applications.
We collected $2,403$ \canvas  issue reports from $123$ open-source GitHub projects that use the \canvas.
We selected a baseline taxonomy based on known aspects of the \canvas, which also provided a set of initial  issue types during our manual analysis. 
We conducted an iterative manual classification process using a random sample of $332$  issue reports.
Through our study, we empirically constructed a taxonomy of testable \canvas  issues that provides the basis for future research on \canvas testing. Our study also provided several insights for developers when working with the \canvas:
\begin{itemize}
	\item \textit{Visual}  issues are the most frequently encountered \canvas  issues in the studied issue reports.
	\item Often,  issues which appear to be caused by the \canvas are in fact due to issues in other parts of the web application.
	\item Browsers do not necessarily provide the same \canvas functionality, and this is particularly true for outdated browsers.
	\item Use of the \canvas brings specific performance concerns in web applications.
	\item JavaScript methods built into browsers for saving \canvas data may provide an unexpected result.
	\item Cross-origin resource sharing policies must be correctly used to load \canvas resources from foreign domains.
\end{itemize}
	
	\section*{Acknowledgments}
	The research reported in this article has been supported by Prodigy Education and the Natural Sciences and Engineering Research Council of Canada under the Alliance Grant project ALLRP 550309.

	\ifCLASSOPTIONcaptionsoff
	\newpage
	\fi
	

	\printbibheading
	
	\balance
	
	\AtNextBibliography{\footnotesize}
	\printbibliography[nottype=misc, title={Works Cited}, heading=subbibliography]
	
	\AtNextBibliography{\footnotesize}
	\printbibliography[type=misc, title={GitHub Issue Reports Cited}, heading=subbibliography]

\end{document}